 \documentclass[final,3p,times,twocolumn,sort&compress]{elsarticle}
 \usepackage{graphicx,amssymb,amsmath,amsthm,lineno,enumitem}
 \usepackage{dcolumn,multirow}
 \usepackage[center]{subfigure}
 \usepackage{threeparttable}
\biboptions{square}

\begin{document}

\begin{frontmatter}



\title{Self-Diffusion of Water through Thermally Activated Membranes}


\author[ufsa]{Carlos Handrey Araujo Ferraz\corref{cor1}}
\ead{handrey@ufersa.edu.br}
\cortext[cor1]{Corresponding author}
\address[ufsa]{{Department of Natural Sciences, Mathematics and Statistics, Universidade Federal Rural do Semi-\'Arido, Mossor\'o 59625-900, RN, Brazil}
}

\begin{abstract}
Diffusion processes involving membranes are of fundamental importance in both science and technology, since membranes serve as selective interfaces that regulate the transport of mass, charge, and information across multiple length and time scales. In this study, we employ molecular dynamics (MD) simulations to calculate the self-diffusion coefficient, tetrahedral order parameter and  hydrogen-bond (HB) lifetime of SPC/E water over a wide high-temperature range in the presence of thermally activated membranes (TAMs). These membranes exhibit thermally controlled stochastic behavior that locally influences particle dynamics by probabilistically inducing elastic scattering events as particles traverse the membrane. The stochastic behavior of the membranes is governed by a sigmoidal profile, which depend on the reduced temperature of the system. The effective activation energy for diffusion is estimated for several membrane configurations. It is found that the diffusion coefficients generally decrease with an increase in the number of membranes and are in reasonable agreement with the Arrhenius approximation at high temperatures. Additionally, both the tetrahedral order parameter and the HB lifetime are only locally sensitive to the action of the membranes.      
\end{abstract}


\end{frontmatter}

%%\linenumbers

\section{Introduction \label{sec:int}}

From a theoretical standpoint, diffusion provides a direct bridge between microscopic dynamics and macroscopic behavior. The classical formulation given by Fick’s laws~\cite{fick1855}, establishes a phenomenological description in which fluxes are proportional to concentration gradients, while more rigorous treatments emerge from statistical mechanics via the random walk model and the Einstein–Smoluchowski relation~\cite{einsten1905}. These frameworks are essential for understanding how molecular-level interactions give rise to emergent transport coefficients such as the diffusion constant. Furthermore, diffusion plays a key role in non-equilibrium thermodynamics, where it contributes to entropy production and irreversible processes~\cite{groot1984}. In materials science, it governs processes such as alloy formation, sintering, and phase transformations. In biology and medicine, diffusion is fundamental to processes like oxygen transport in tissues~\cite{hess2024}, drug delivery~\cite{he2024, mohanto2025}, and cellular signaling~\cite{soltan2022, yuan2024}.

In particular, diffusion processes involving membranes are of fundamental importance in both science and technology because membranes act as selective barriers that regulate the transport of mass, charge, and information at multiple scales. In biological systems, membrane-mediated diffusion governs essential processes such as ion transport, nutrient uptake, and signal transduction, thereby underpinning cellular homeostasis and physiological function ~\cite{albert2022, phillips2013}. From a physical point of view, they are characterized by spatial heterogeneity, confinement, and selective permeability that often result in deviations from classical Fickian diffusion and necessitate more sophisticated frameworks incorporating facilitated transport, active processes, and coupling with electrochemical gradients ~\cite{hille2001, keener2009}. In technological settings, membrane diffusion is key to desalination, gas separation, energy storage (e.g., fuel cells and batteries), and controlled drug delivery applications, where transport efficiency and selectivity often limit performance ~\cite{baker2012, strathmann2011}.

However, important issues have not yet been understood, particularly at the level of molecular dynamics, where the relationship between membrane structure and permeability is not yet fully understood. The contribution of membrane heterogeneity—including dynamic structural fluctuations and the presence of defects—remains insufficiently characterized, especially in biological and soft-matter systems. Furthermore, theoretical descriptions are complicated by non-equilibrium effects, such as those arising in active membranes and driven transport processes, which challenge the development of unified and predictive frameworks. In particular, problems involving active membranes with stochastic behavior conditioned by factors such as temperature and pressure introduce additional complexity (e.g., jump processes) into the dynamics of interacting particles. Addressing these questions is of essential importance for advancing both fundamental science and practical applications in membrane technology.

In parallel, water, despite its apparent molecular simplicity, exhibits remarkably complex dynamical behavior in its diffusion across different temperature regimes. The self-diffusion coefficient, which quantifies the rate at which water molecules move randomly through the liquid medium, not only increases with temperature but does so in a non-linear and anomalous manner compared to simple liquids~\cite{holz2000, pettersson2016}. This transport property plays a central role in a wide range of biological, chemical, and physical processes, from cellular homeostasis to enzymatic activity and membrane separation phenomena~\cite{laage2011, laage2012}.

One of the most persistent and well-studied topics is the precise characterization of how the diffusion coefficient varies with temperature and why this behavior deviates from the simple Arrhenius model. From low to high temperatures, the diffusion behavior of water exhibits a clear transition between distinct dynamical regimes. In the supercooled region ($\sim 180–273 \, \mathrm{K}$), water shows strongly non-Arrhenius behavior, characterized by a rapid slowdown of diffusion due to the formation of a highly structured and long-lived hydrogen-bond network, leading to cooperative and heterogeneous molecular dynamics~\cite{debenedetti2003, angell2008, galamba2016}. As the temperature increases toward ambient conditions ($\sim \, 280–320 \mathrm{K}$), the system displays an approximately Arrhenius-like behavior over a limited range, where the effective activation energy for diffusion remains nearly constant and reflects a balance between bond breaking and reformation~\cite{holz2000, mills1973}. At higher temperatures (above $\sim 320 \,\mathrm {K} $), thermal fluctuations progressively disrupt the hydrogen-bond network, reducing structural correlations and leading to a crossover toward simpler, near-Arrhenius dynamics, typical of ordinary liquids~\cite{kumar2006, gallo2016, gomez2022}.

\begin{figure}[!t]
	\centering
	\includegraphics*[scale=0.35,angle=0]{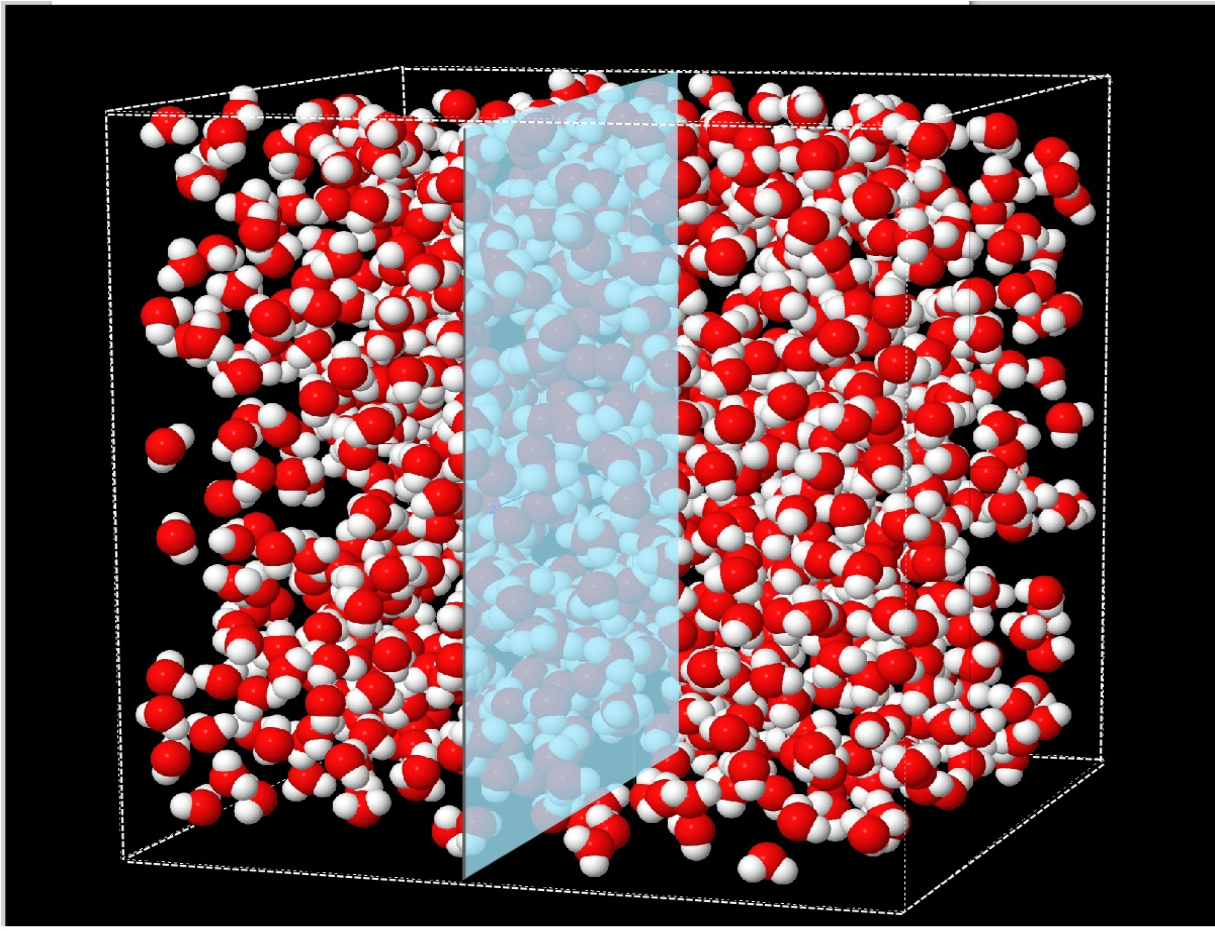}
	\caption{ The schematic depicts a three-dimensional simulation box with periodic boundary conditions containing $N=10^3$  water molecules. A single thermoactivated membrane (rendered in transparent blue) is positioned at the center of the box.}\label{fig:01}
\end{figure}

\begin{figure}[!t]
	\centering
	\includegraphics*[scale=0.40,angle=0]{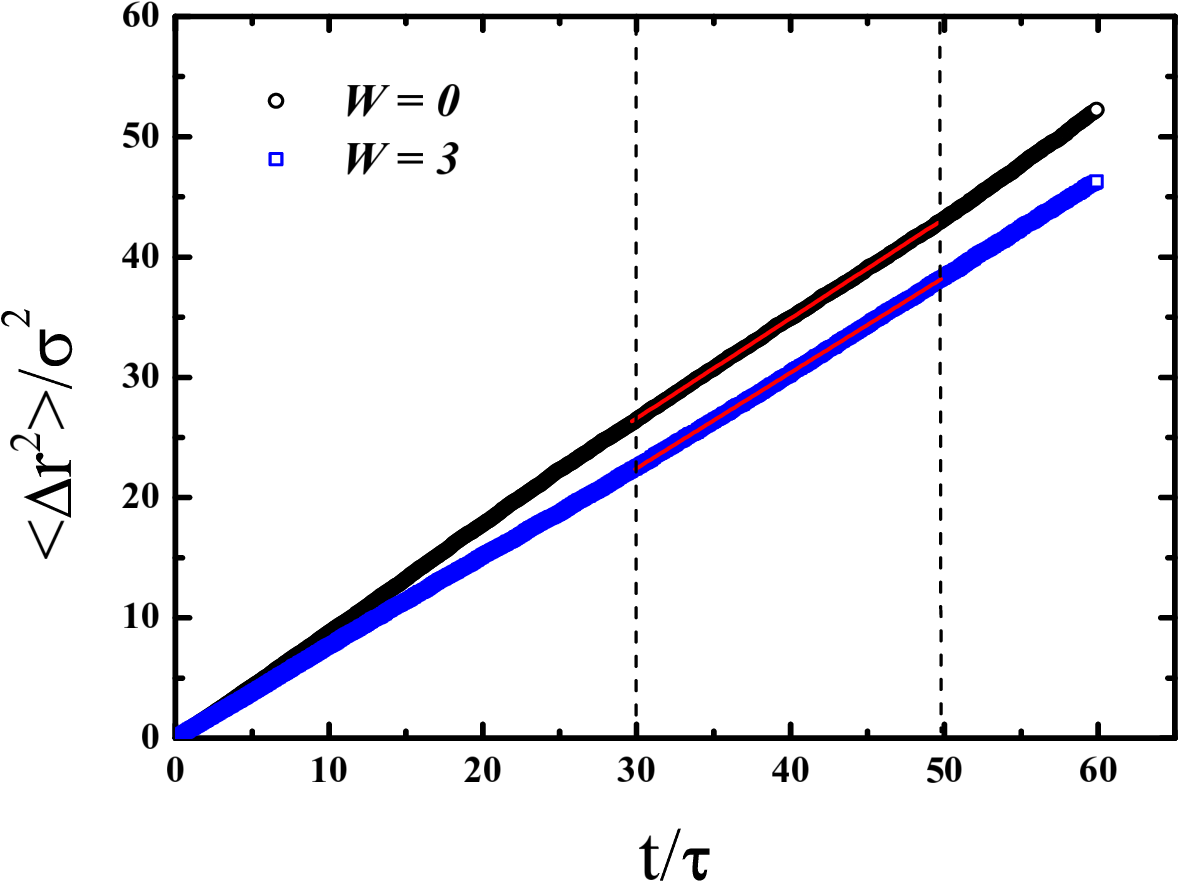}
	\caption{Plot of the MSD as a function of time at temperature $T = 387.3\,\mathrm{K}$ and density $\rho = 1.0\,\mathrm{g/cm^3}$ for two different $W$-membrane configurations (see Sec.~\ref{sec:md}). The self-diffusion coefficient, $D$, is determined from the slope of the MSD according to Eq.~\eqref{eq:18}. The asymptotic diffusion coefficient is obtained by performing a least-squares linear fit to the long-time regime (red solid lines), over the time interval from $30\tau$ to $50\tau$ (black dashed lines). Similar plots are obtained for the other temperatures and densities investigated. The scaling parameters $\sigma$ and $\tau$ are listed in Table~\ref{table:01}.
      }\label{fig:02}
\end{figure}

\begin{figure*}[!t]
	\centering
	\begin{minipage}[t]{1.0\linewidth}
		\centering
		%\subfigcapskip = 10pt
		\subfigure[Case $W=0$]{\label{fig:03a}\includegraphics[scale=0.38, angle=0]{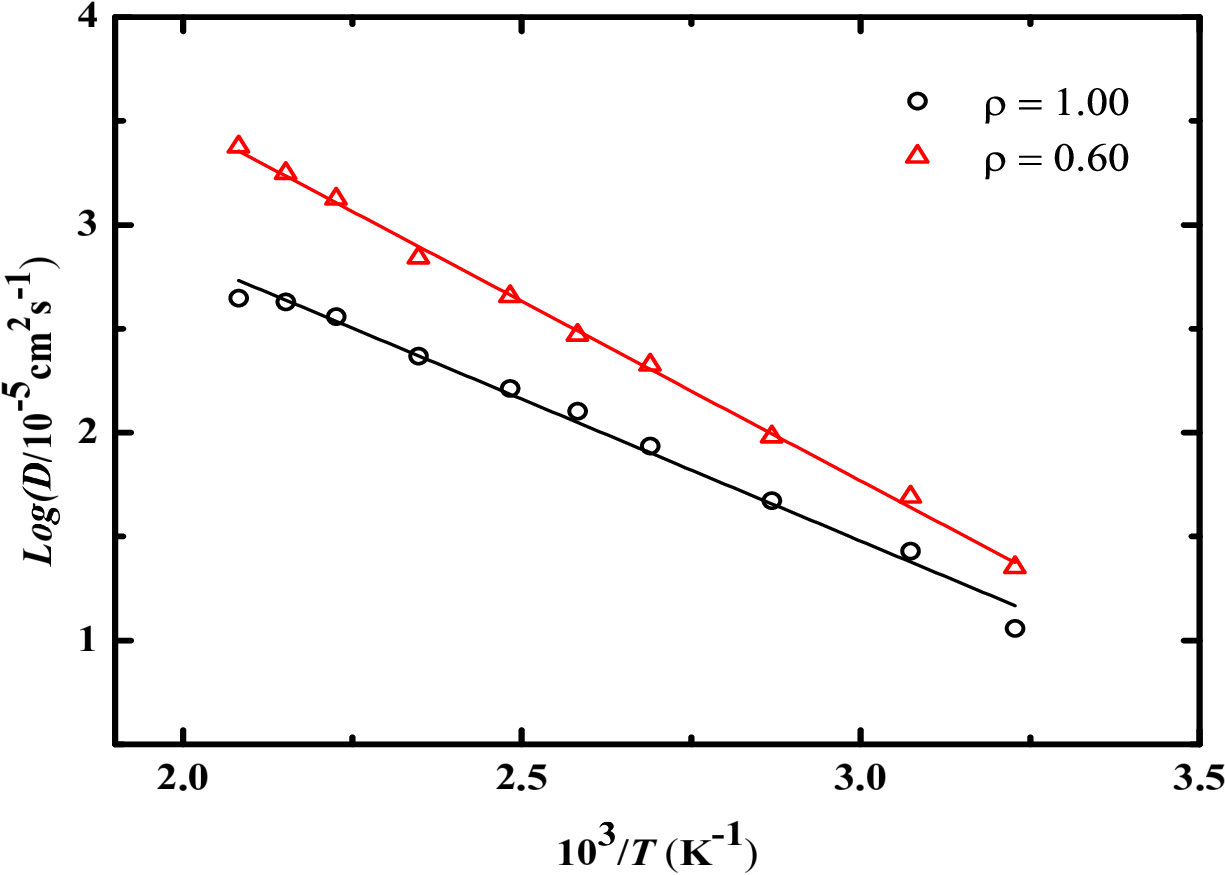}}
		%\hspace{5cm}
		\subfigure[Case $W=1$]{\label{fig:03b}\includegraphics[scale=0.38, angle=0]{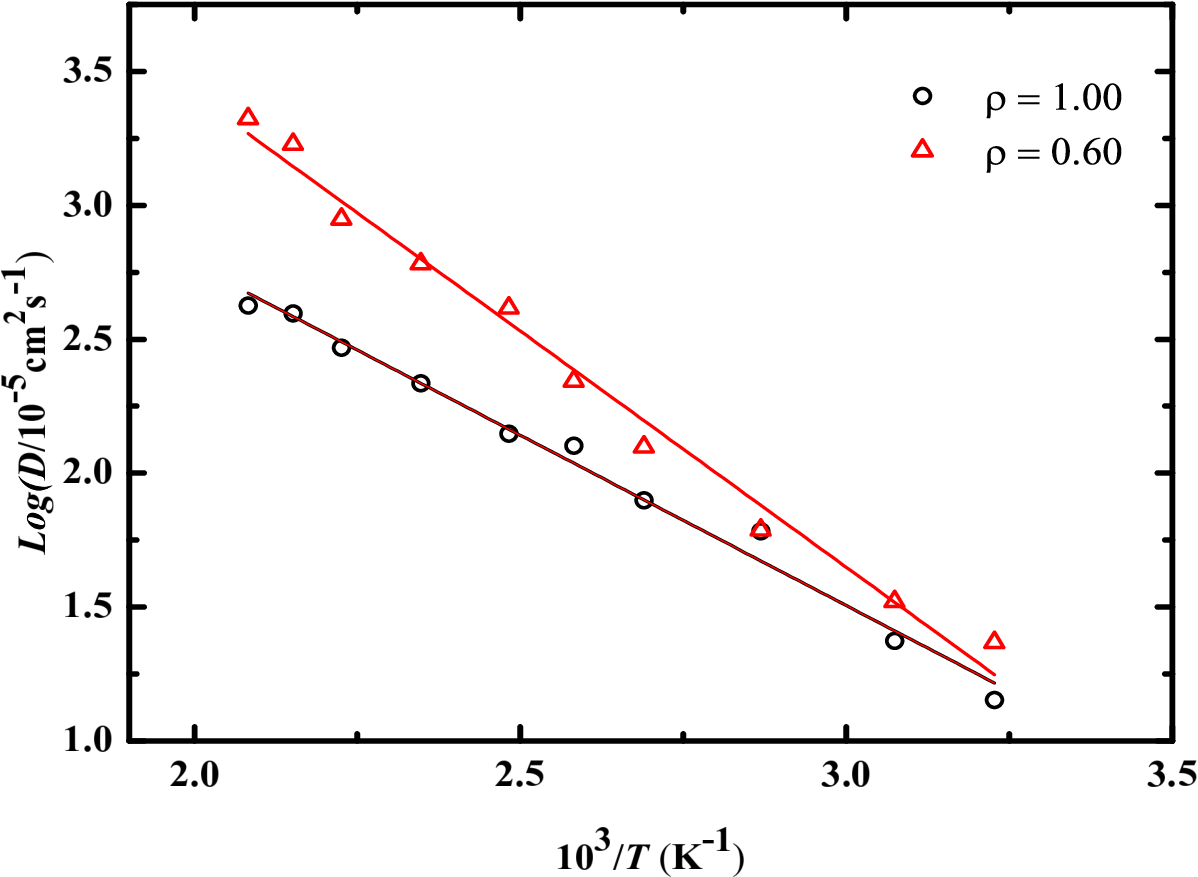}} 
		%\qquad \qquad
		\subfigure[Case $W=2$]{\label{fig:03c}\includegraphics[scale=0.38, angle=0]{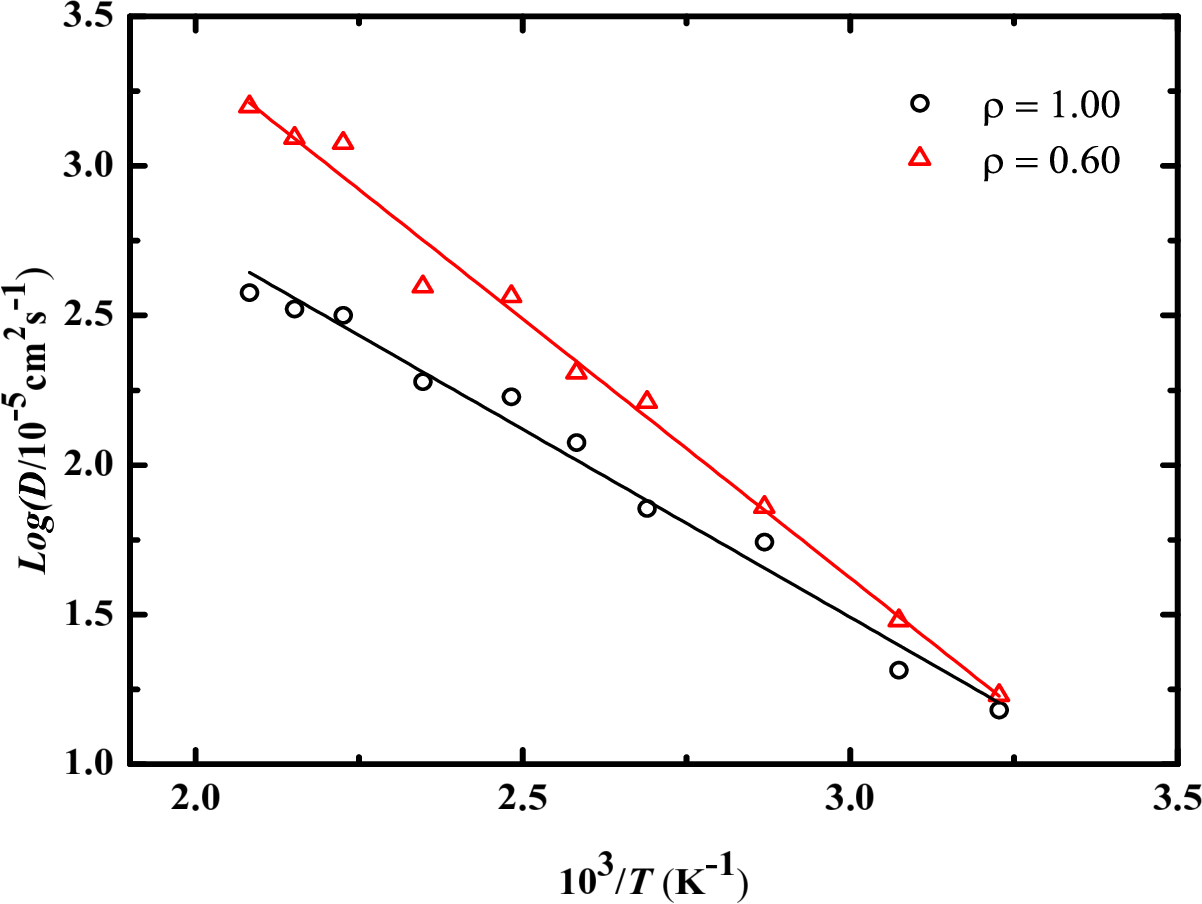}}
		% \hspace{5cm}
		\subfigure[Case $W=3$]{\label{fig:03d}\includegraphics[scale=0.38, angle=0]{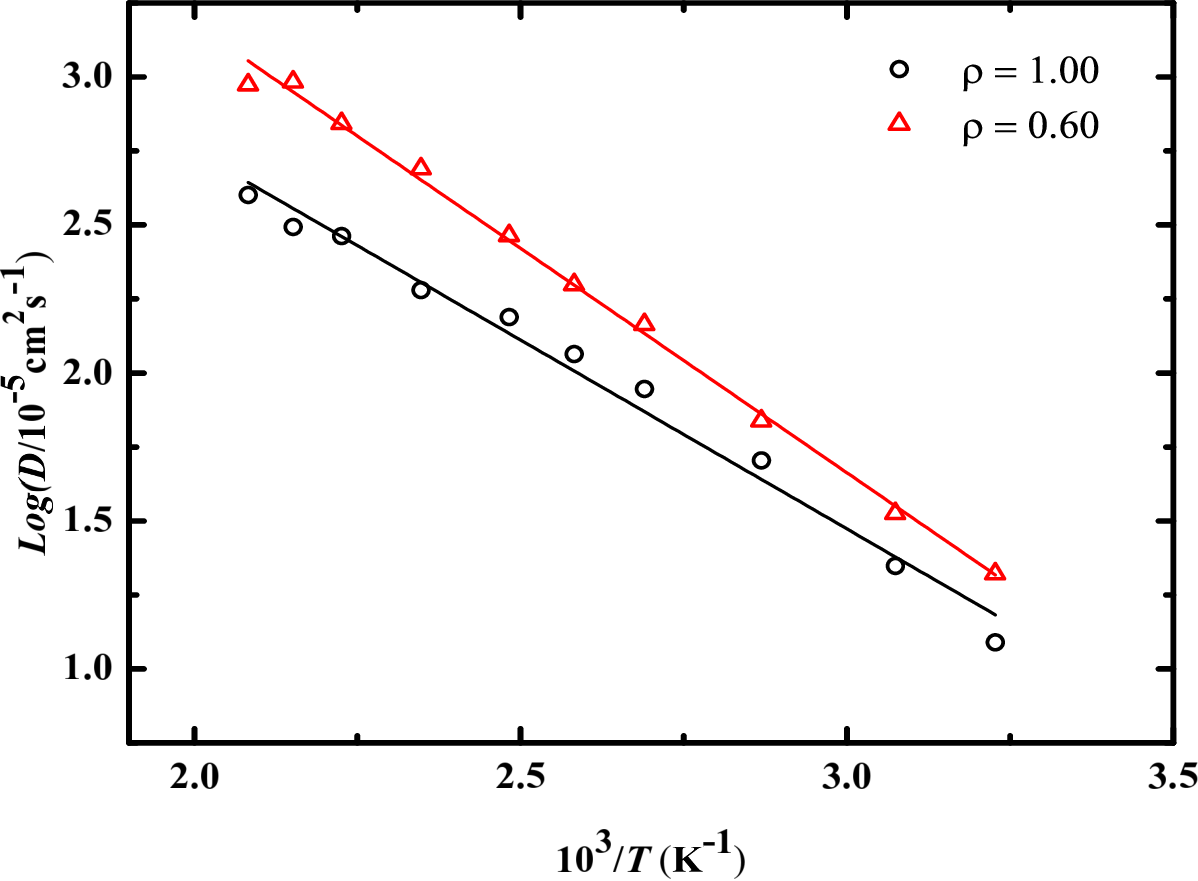}}
	\end{minipage}
	\caption{Semi-log plot of the self-diffusion coefficients (in physical units) as a function of temperature for the different TAM configurations and the two water-density regimes. The Arrhenius curve fitted to data are also shown (solid lines) and the respective activation energies ($E_{a}$) and pre-exponential factor ($D_{0}$) are reported in Table \ref{table:02}. Error bars are of the same size of plot symbols and are not showed. (a) Absence of TAMs ($W=0$). (b) Presence of a single TAM ($W=1$). (c) Presence of two TAMs ($W=2$). (d) Presence of three TAMs ($W=3$).} \label{fig:03}
\end{figure*}

\begin{figure*}[!t]
	\centering
	\begin{minipage}[t]{1.0\linewidth}
		\centering
		%\subfigcapskip = 10pt
		\subfigure[$T=310\, \mathrm{K}$]{\label{fig:04a}\includegraphics[scale=0.40, angle=0]{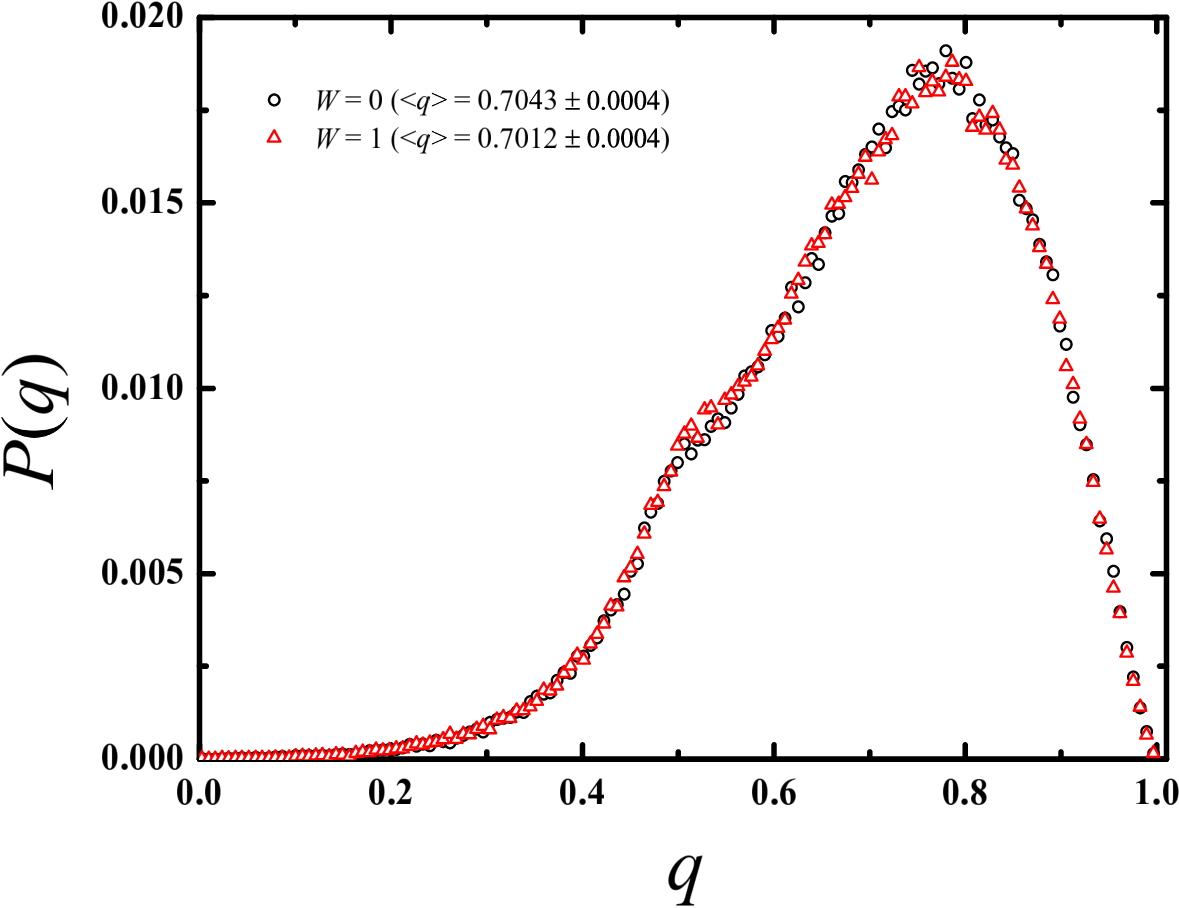}}
		%\hspace{5cm}
		\subfigure[$T=387\, \mathrm{K}$]{\label{fig:04b}\includegraphics[scale=0.40, angle=0]{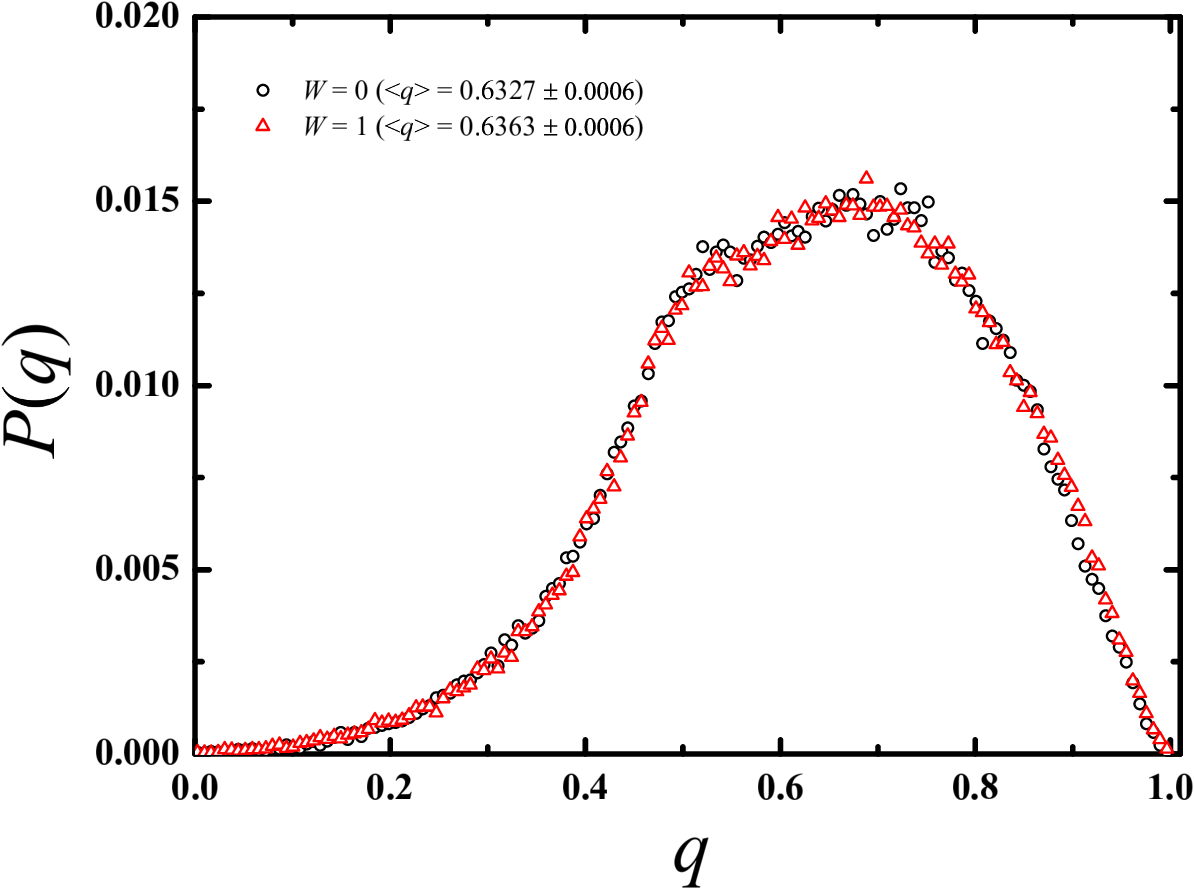}} 
	\end{minipage}
	\caption{ Tetrahedrality distribution, $P(q)$, and the mean tetrahedrality, $\langle q \rangle$ (see the plot legends), at temperatures (a) $T=310,\mathrm{K}$  and (b) $T=387,\mathrm{K}$ for the two $W$ cases considered, at a water density of $1.0\,\mathrm{g/cm^3}$. According to Eq.~\eqref{eq:13}, at $T=310\,\mathrm{K}$, the probability that a given water molecule undergoes elastic scattering at the membrane is $p\simeq 0.52$, whereas at $T=387\,\mathrm{K}$, this probability increases to $p\simeq 0.75$.} \label{fig:04}
\end{figure*}

\begin{table}[!b]
	\setlength{\tabcolsep}{10pt}
	\small
	\centering
	\begin{threeparttable}
		\caption{\label{table:01} Physical units used in the simulations.}
		\begin{tabular}{lcccc}
			\hline \hline \\
			Units & & & & Value \\ \hline \\
			Length ($\sigma$) & & & & $3.166 \, \mathrm{\AA}$ \\
			Time ($\tau$)    & & & &$1.673 \times 10^{-12} \, \mathrm{s}$ \\
			Energy ($\varepsilon$) & & & & $0.644 \, \mathrm{kJ\,mol^{-1}}$ \\
			Temperature ($T_{0}$)  & & & & $77.457\, \mathrm{K}$  \\
			\hline \hline
		\end{tabular}
	\end{threeparttable}
\end{table}

In this study, we performed molecular dynamics (MD) simulations to investigate the self-diffusion of water and HB dynamics over a wide temperature range in the presence of thermally activated membranes (TAMs). These membranes exhibit thermally controlled stochastic behavior that locally affects particle dynamics by randomly reversing the transverse velocity component of molecules ($v \rightarrow -v$) as they pass through the membrane. The system temperature is maintained constant using a thermostat~\cite{Bussi2007}, while the pressure is allowed to fluctuate. The stochastic behavior of the membranes is described by a sigmoidal probability function, and several membrane configurations are considered.
The SPC/E~\cite{berendsen1987} potential was used to model water along with periodic boundary conditions. We chose this model mainly because it provide a good balance between computational cost and accuracy, making it particularly appealing when the goal is to investigate physical trends, transport mechanisms (such as diffusion), or confinement effects (such as in the case of membranes), without the computational burden of more complex models. The effective activation energy is estimated over a wide high-temperature range ($\sim 310\text{--}465\,\mathrm{K}$) for multiple membrane configurations. The main objectives of this study are twofold: (i) to characterize and understand how particle diffusion is influenced by TAMs, and (ii) to investigate whether the stochastic action of the membranes modifies the hydrogen-bond (HB) network and the HB lifetime of the system.

\begin{figure*}[!t]
	\centering
	\begin{minipage}[t]{1.0\linewidth}
		\centering
		%\subfigcapskip = 10pt
		\subfigure[$T=310\, \mathrm{K}$]{\label{fig:05a}\includegraphics[scale=0.40, angle=0]{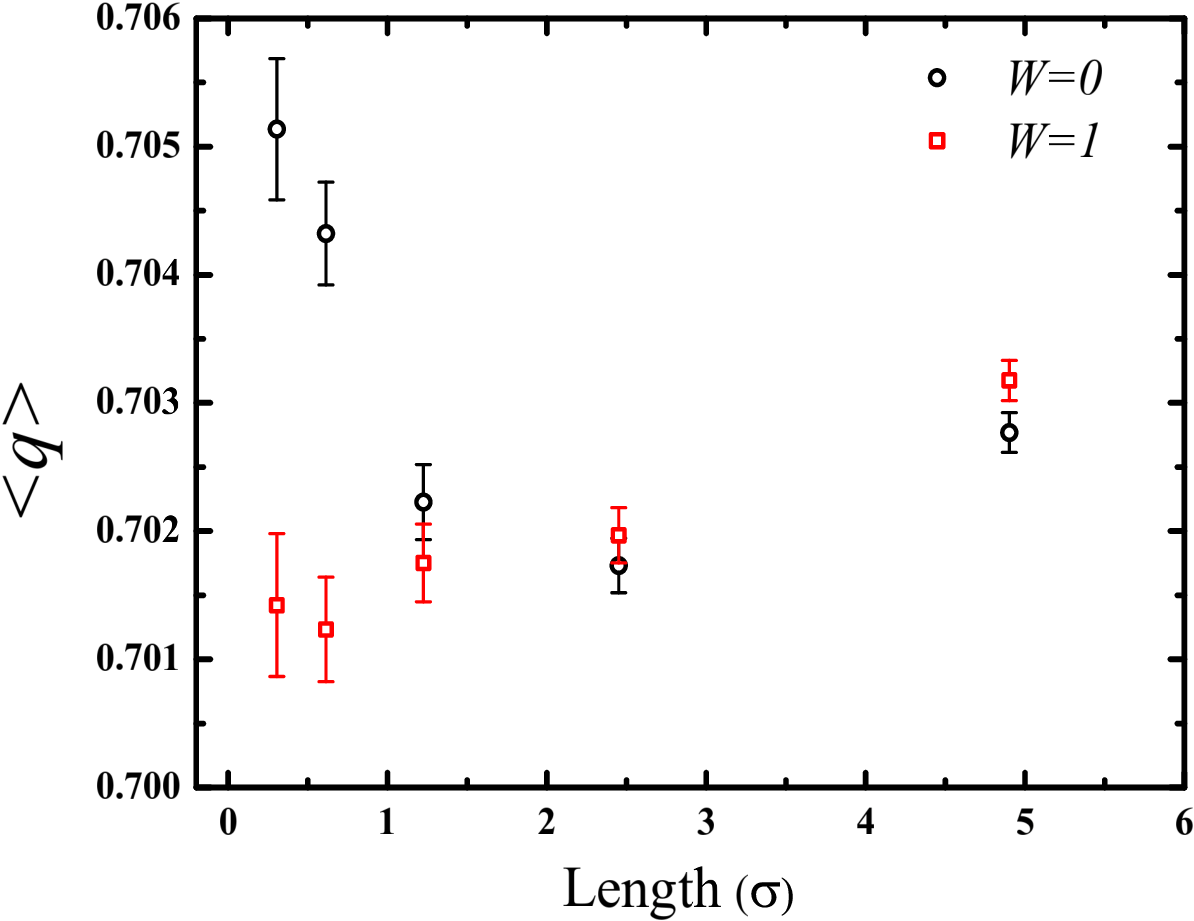}}
		%\hspace{5cm}
		\subfigure[$T=387\, \mathrm{K}$]{\label{fig:05b}\includegraphics[scale=0.40, angle=0]{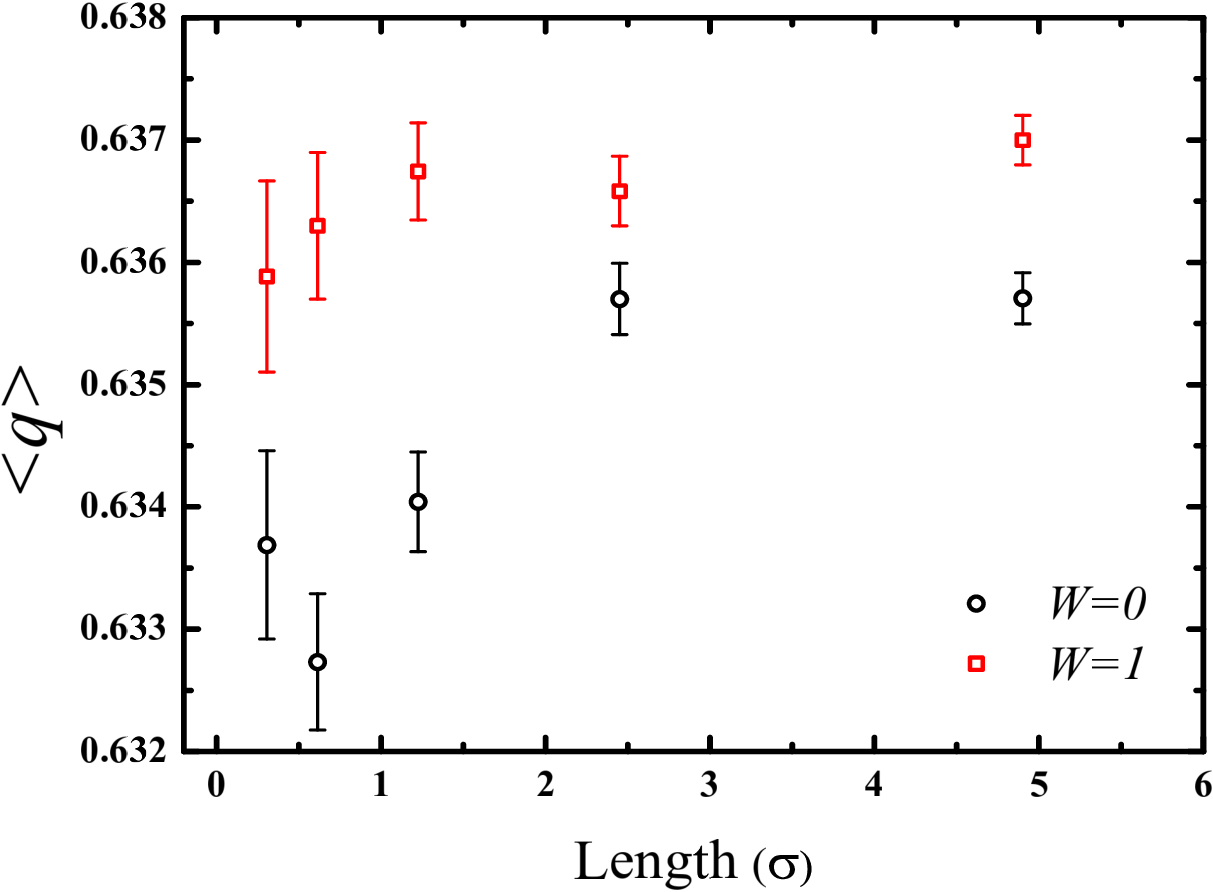}} 
	\end{minipage}
\caption{Mean tetrahedrality, $\langle q \rangle$, calculated for water molecules within virtual boxes of different lengths centered at the membrane ($x=0$), at (a) $T=310,\mathrm{K}$ and (b) $T=387,\mathrm{K}$, for a water density of $1.0\,\mathrm{g/cm^3}$. Results are shown for two different $W$ cases, with the virtual-box length ranging from $L/32$ to $L/2$ ($L=9.8\, \sigma$).} \label{fig:05}
\end{figure*}

The manuscript is organized as follows. In section~\ref{sec:mms}, the theoretical background of the model is presented. Section~\ref{sec:md} provides the details of the MD simulations. In section~\ref{sec:r}, we present and discuss the results. Finally, in section~\ref{sec:c}, we summarize our conclusions.
 
 \section{Theoretical Background \label{sec:mms}}
 
 We consider a system of $N$ interacting molecules confined in a cubic domain $\Omega = [-L/2,L/2)^3$ with periodic boundary conditions. Let $\mathbf{r}_i$ and $\mathbf{v}_i$ denote the position and velocity of the center of mass of the $i$-th molecule, respectively. Each molecule consists of three interaction sites labeled by $\alpha=\{1,2,3\}$, located at positions $\mathbf{r}_{i\alpha}$, with partial charges $q_{i\alpha}$. For water-like models, these three sites correspond to one oxygen atom (O) and two hydrogen atoms (H), with the molecular geometry constrained by rigidity.
 
In addition to periodic boundaries, the system may contain planar thermoactivated membranes (see Fig.~\ref{fig:01}) located at
 \begin{equation} \label{eq:1}
 	x = x_w, \qquad -L/2 < x_w < L/2,
 \end{equation}
 with unit normal $\mathbf{n} = \hat{\mathbf{e}}_x$.
 
 \subsection{Interparticle interactions}
 
 The total potential energy is given by
 \begin{equation}  \label{eq:2}
 	U(\mathbf{r}^N) = \sum_{1 \le i < j \le N}
 	\left[U_{ij}^{C} + 	U_{ij}^{VW}
 	\right],
 \end{equation}
 where
 \begin{equation}  \label{eq:3}
 	U_{ij}^{C} =
 	\sideset{}{'}\sum_{\mathbf{n} \in \mathbb{Z}^3}\sum_{\alpha \in i} \sum_{\beta \in j}
 	\dfrac{q_{i\alpha} q_{j\beta}}{\left| \mathbf{r}_{i\alpha, j \beta} + L \mathbf{n} \right|},
 \end{equation}
 denotes the Coulomb potential under periodic boundary conditions (Ewald summation). The first sum is over all integer vectors $\mathbf{n}$, and the prime indicates omission of the self-interaction term $i=j$, $\alpha=\beta$ when $\mathbf n= 0$.  
 
 While van der Waals interaction is modeled using a Lennard--Jones potential~\cite{Ferraz2023, Ferraz2021}:
 \begin{equation}  \label{eq:4}
 	U_{ij}^{VW} = \sum_{\alpha \in i} \sum_{\beta \in j} \dfrac{A_{\alpha \beta}}{{r}_{i\alpha, j\beta}^{12}}-\dfrac{B_{\alpha \beta}}{{r}_{i\alpha, j\beta}^{6}},
 \end{equation}
 being $A_{\alpha \beta}$ and $B_{\alpha \beta}$ model parameters~\cite{berendsen1987} and $\mathbf r_{i\alpha, j\beta}=\mathbf r_{i\alpha}-\mathbf r_{j\beta}$ in above equations.
 
 \subsection{Equations of motion}
 
Outside the membrane, the time evolution of the system is governed by the equations of motion:

 \begin{equation}  \label{eq:5}
 	m_i \dot{\mathbf{v}}_i = \mathbf{F}_i
 \end{equation}
 and 
 \begin{equation}  \label{eq:6}
 	{ \mathbb I}_{i}\dot{\pmb{\omega}}_i + {\pmb {\omega}}_i\times ({\mathbb I}_{i}\pmb{\omega}_i)=\pmb{\tau}_{i},
 \end{equation}
 
 where $m_i$ and ${ \mathbb I}_{i}$ are the mass and inertia tensor (expressed in its principal axes) of the $i$-th particle, respectively, and ${\mathbf{v}}_i$ and ${\pmb {\omega}}_i$ denote its linear and angular velocities. The vectors $\mathbf{F}_i$ and $\pmb{\tau}_{i}$ are given by
  
 \begin{equation}  \label{eq:7}
 	\mathbf{F}_i = \sum_{\alpha \in i} \mathbf{f}_{i\alpha} = -\nabla_i U(\mathbf{r}^N)	
 \end{equation}
 and
 \begin{equation} \label{eq:8}
 	\pmb{\tau}_{i}= \sum_{\alpha \in i} (\mathbf{r}_{i\alpha}-\mathbf{r}_i) \times \mathbf{f}_{i\alpha},
 \end{equation}
 
  and represent the total force and torque acting on particle $i$, respectively.

 \subsection{Membrane dynamics}
 
 \begin{figure*}[!t]
 	\centering
 	\begin{minipage}[t]{1.0\linewidth}
 		\centering
 		%\subfigcapskip = 10pt
 		\subfigure[Case $W=0$]{\label{fig:06a}\includegraphics[scale=0.39, angle=0]{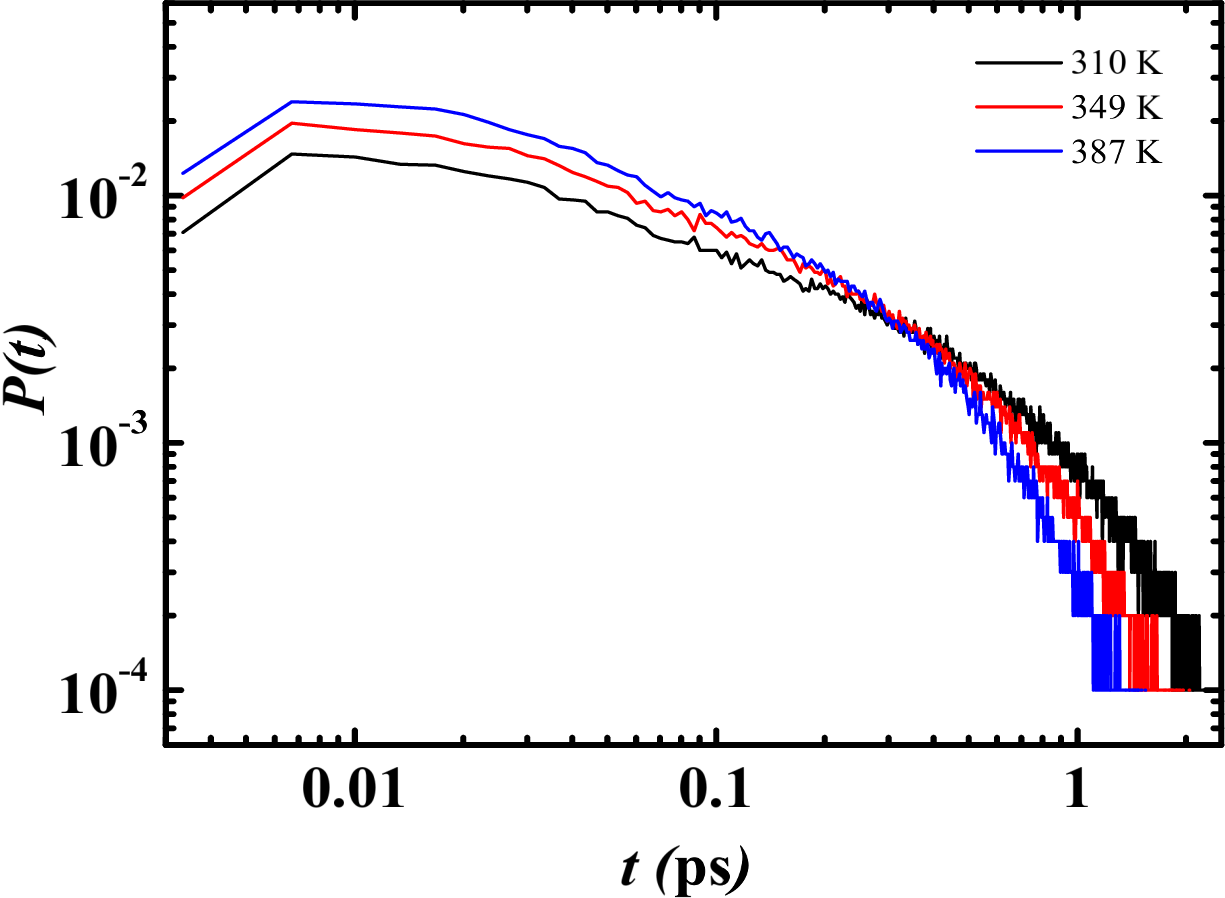}}
 		%\hspace{5cm}
 		\subfigure[Case $W=3$]{\label{fig:06b}\includegraphics[scale=0.39, angle=0]{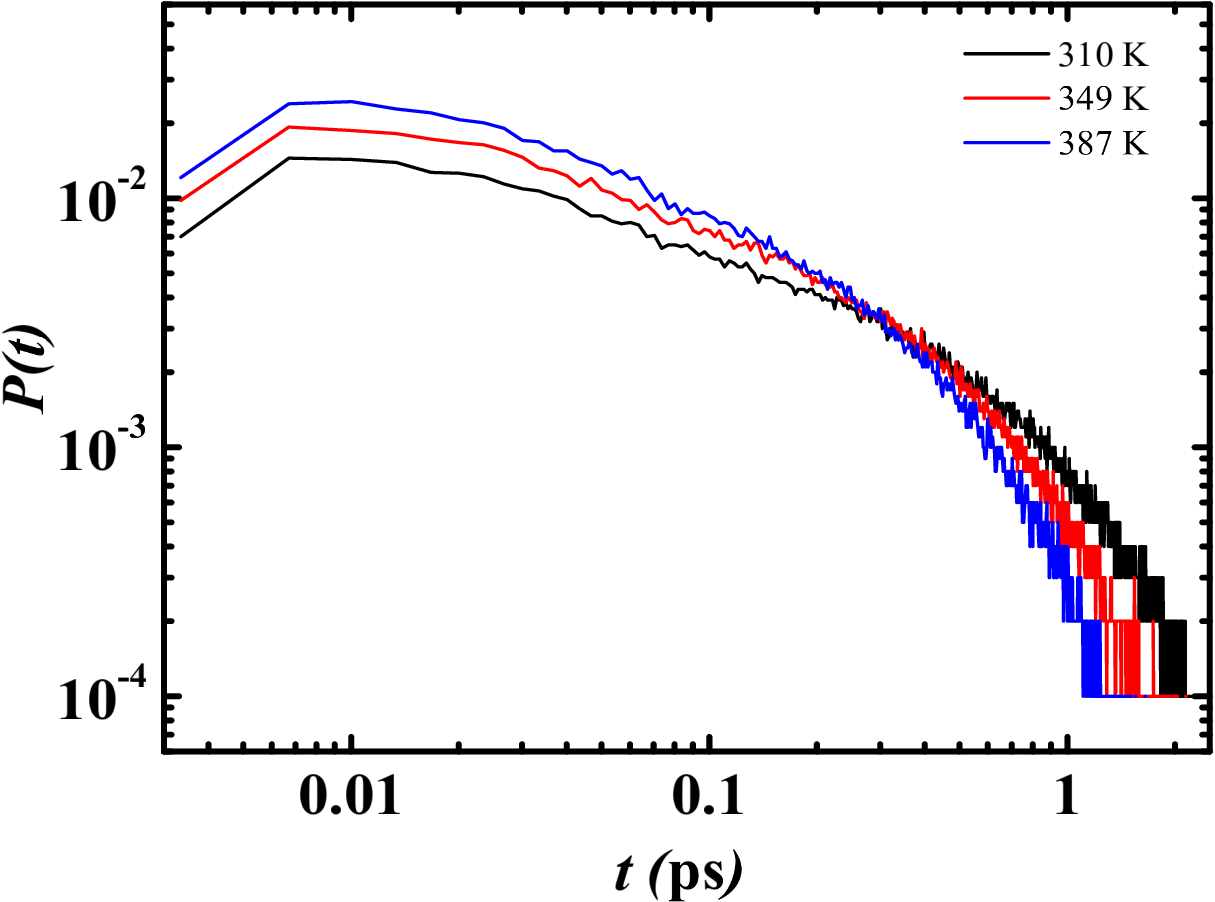}} 
 	\end{minipage}
 \caption{Log--log plots of the HB lifetime probability distributions at three different temperatures for (a) $W=0$ and (b) $W=3$. Typical exponential tails are observed in all cases considered.}
  \label{fig:06}
 \end{figure*}
 
 When a particle $i$ reaches a membrane at $x_w$, its center-of-mass velocity is updated according to a stochastic rule. Let $\mathbf{v}_i^-$ and $\mathbf{v}_i^+$ denote the velocities immediately before and after interaction with the membrane. Then
 \begin{equation} \label{eq:9}
 	\mathbf{v}_i^+ =
 	\begin{cases}
 		\mathcal{R}\mathbf{v}_i^-, & \text{with probability } p, \\
 		\mathbf{v}_i^-, & \text{with probability } 1 - p,
 	\end{cases}
 \end{equation}
 where $\mathcal{R}$ is the reflection operator defined by
 \begin{equation} \label{eq:10}
 	\mathcal{R}\mathbf{v}
 	=
 	\mathbf{v} - 2(\mathbf{v} \cdot \mathbf{n}) \mathbf{n}.
 \end{equation}
 For the present geometry, this reduces to
 \begin{equation} \label{eq:11}
 	(v_x, v_y, v_z) \rightarrow (-v_x, v_y, v_z).
 \end{equation}
 The particle position remains continuous across the membrane:
 \begin{equation} \label{eq:12}
 	\mathbf{r}_i^+ = \mathbf{r}_i^-.
 \end{equation}
 In foregoing Eq.~\eqref{eq:9}, the probability $p$ characterizes the thermally controlled membrane behavior for it obeys a sigmoidal perfil given by
 
 \begin{equation} \label{eq:13}
 p = \frac{1}{1 + b \,\exp\!\left[-(a \,\theta)\right]},
 \end{equation}
where $\theta= T/T_{0}$ is the dimensionless temperature of the system ($T_{0}$ is defined in Table \ref{table:01}), and $a = 1$ and $b = 50$ are the steepness and shift parameters, respectively.
 
 The sigmoidal profile was chosen because it can model systems exhibiting gradual transitions or thresholds, such as phase transitions and reaction kinetics~\cite{Ferraz2018,Ferraz2022}. Its $S$-shaped form allows for the representation of processes that start slowly, accelerate through a midpoint, and then level off, making them useful in understanding stability, feedback mechanisms, and non-linear responses in complex systems. 
 
 \subsection{Kinetic description}
 
 Let $\rho(\mathbf{r}^N, \mathbf{v}^N, t)$ denote the $N$-particle phase-space density. Its evolution is governed by a Liouville equation with a jump operator:
 \begin{equation} \label{eq:14}
 	\frac{\partial \rho}{\partial t}
 	+
 	\sum_{i=1}^N
 	\left[
 	\mathbf{v}_i \cdot \nabla_{\mathbf{r}_i} \rho
 	+
 	\frac{\mathbf{F}_i}{m_i} \cdot \nabla_{\mathbf{v}_i} \rho
 	\right]
 	=
 	\sum_{i=1}^N \mathcal{J}_i[\rho],
 \end{equation}
 where
 \begin{equation} \label{eq:15}
 	\mathcal{J}_i[\rho]
 	=
 	p \, |v_{ix}| \, \delta(x_i - x_w)
 	\left[
 	\rho(\mathbf{r}^N, \ldots, \mathcal{R}\mathbf{v}_i, \ldots, t)
 	-
 	\rho(\mathbf{r}^N, \mathbf{v}^N, t)
 	\right].
 \end{equation}
  
 In a reduced description, the one-particle distribution function $f(\mathbf{r}, \mathbf{v}, t)$ satisfies
 \begin{equation} \label{eq:16}
 	\begin{split}
 	\frac{\partial f}{\partial t}
 	& +
 	\mathbf{v} \cdot \nabla_{\mathbf{r}} f
 	+
 	\frac{\mathbf{F}(\mathbf{r},t)}{m} \cdot \nabla_{\mathbf{v}} f
 	=     \\
   & p \, |v_x| \, \delta(x - x_w)
 	\left[
 	f(\mathbf{r}, \mathcal{R}\mathbf{v}, t)
 	-
 	f(\mathbf{r}, \mathbf{v}, t)
 	\right].
 	\end{split}
 \end{equation}
 
 Explicitly,
 \begin{equation} \label{eq:17}
 	f(\mathbf{r}, \mathcal{R}\mathbf{v}, t)
 	=
 	f(x,y,z,-v_x,v_y,v_z,t).
 \end{equation}

\section{\label{sec:md} MD simulations}

\begin{table*}[ht]
	\setlength{\tabcolsep}{10pt}
	\small
	\centering
	\caption{\label{table:02} Activation energy $E_a$ and pre-exponential factor $D_0$ 
		for different water densities and TAM configurations.}
	
	\begin{tabular}{c c c c c c}
		\hline \hline \\
		$\rho$ ($\mathrm{g/cm^3}$) & Parameter 
		& $W=0$ & $W=1$ & $W=2$ & $W=3$ \\\\
		\hline 
		
		\multirow{2}{*}{$1.00$}
		& $E_a$ ($\mathrm{kJ/mol}$) & $11.36 \pm 0.42$ & $10.58 \pm 0.42$
		& $10.45 \pm 0.50$ & $10.61 \pm 0.42$ \\
		
		& $D_0$ ($\times 10^{-5}\, \mathrm{cm^2/s}$) & $265 \pm 34$ & $204 \pm 24$
		& $192 \pm 29$ & $200 \pm 28$ \\
		\hline
		
		\multirow{2}{*}{$0.60$}
		& $E_a$ ($\mathrm{kJ/mol}$)& $14.37 \pm 0.25$ & $14.67 \pm 0.58$
		& $14.40 \pm 0.50$ & $12.60 \pm 0.25$ \\
		
		& $D_0$ ($\times 10^{-5}\, \mathrm{cm^2/s}$) & $1043 \pm 73$ & $1032 \pm 186$
		& $916 \pm 147$ & $498 \pm 40$ \\
		
		\hline \hline
	\end{tabular}
\end{table*}

The MD simulations were performed to calculate the self-diffusion coefficients ($D$), tetrahedral order parameter ($q$) and HB lifetime ($\tau_{HB}$) of SPC/E water. The translational equations of motion (Eq.~\eqref{eq:5}) were numerically solved using the standard leapfrog method~\cite{Hockney1970}, while the rotational equations (Eq.~\eqref{eq:6}) were handled using an implicit leapfrog scheme in the quaternion representation~\cite{Omelyan1998}. A canonical velocity-rescaling thermostat~\cite{Bussi2007} was employed during the simulations to maintain the system temperature constant. A total of $N = 10^3$ molecules were confined in a cubic box with periodic boundary conditions. The production runs were carried out for $9 \times 10^4$ MD time steps in the NVT ensemble, after an equilibration period of $10^4$ MD time steps for the purpose of diffusion calculations. The simulations were performed with a time step of approximately $1.7 \,\mathrm{fs}$. \newline

The self-diffusion coefficients were calculated through the Einstein expression~\cite{PathriaBeale2011}
\begin{equation} \label{eq:18}
	D = \lim_{t\rightarrow \infty}\dfrac{1}{6Nt} \biggl \langle \sum_{i=1}^{N}\biggl [\mathbf{r}_{i}(t)-\mathbf{r}_{i}(0)\biggr ]^2 \biggr \rangle,
\end{equation}
where $t$ is the time-lag and $<\cdots>$ denotes an average over a sufficiently large number of independent samples. Data were collected every 100 time steps, yielding 20 independent datasets over the course of the simulation. Each dataset contains 600 independent samples obtained at different time lags. Averages were then computed over these datasets, resulting in 600 estimates of the diffusion coefficient $D$ at different time lags. This yields quasi-stationary estimates for $D$ at different  time-lags. The asymptotic diffusion coefficients are then obtained from Eq.~\eqref{eq:18} by calculating the slope of the mean-square molecular displacement (MSD) in the sufficiently long-time regime~\cite{Baba2022}. Figure~\ref{fig:02} shows the MSD as a function of time at temperature $T = 387.3\,\mathrm{K}$ and density $\rho = 1.0\,\mathrm{g/cm^3}$ for the different membrane configurations considered (explained below). Similar plots are obtained for other temperatures and densities.

The tetrahedral structure (tetrahedrality) of water was characterized using the orientational order parameter $q$ \cite{Chau1998}, which, in its rescaled form \cite{Errington2001}, is defined as
\begin{equation}  \label{eq:19}
	q=1-\frac{3}{8}\sum_{i=1}^{3}\sum_{j=i+1}^{4}(\cos \theta_{ij}+\frac{1}{3})^2,
\end{equation}
where $\theta_{ij}$ is the angle between the vectors connecting the oxygen atom of a given water molecule to its nearest neighbors, $i$ and $j$. The average value of $q$ varies between 0 (ideal gas) and 1 (perfect tetrahedral HB network).

We investigate the HB dynamics by considering the geometric definition of intact HB, namely, the distance between the donor and acceptor oxygen atoms $r_\mathrm{OO}$ is less than $3.5 \mathrm{\AA}$ and the angle between the intra-molecular O--H bond and $r_\mathrm{OO}$, $\phi_\mathrm{HOO}$, is less than $30^\circ$ \cite{Luzar1996}. The mean HB lifetime is then calculated by 
\begin{equation}  \label{eq:20}
	\tau_\mathrm{HB} = \int_{0}^{\infty} t \, P(t) \,dt,
\end{equation} 
where $P(t)$ is the HB lifetime probability distribution function \cite{Starr1999}, which represents the probability that an initially bonded donor--acceptor pair remains continuously bonded up to time $t$ and breaks at time $t$. The function $P(t)$ is obtained from the simulations by constructing a histogram of the HB lifetimes over all sampled configurations. It is worth noting that this approach does not account for hydrogen bonds that are transiently broken and subsequently reformed within the sampling interval, $\Delta t$. In general, the calculated $\tau_{\mathrm{HB}}$ values depend on the sampling interval, as discussed in Ref.~\cite{Martiniano2013}. In the present simulations, the sampling interval was set to $\Delta t = 3.4\,\mathrm{fs}$.

The interaction of particles with the thermally controlled membrane is implemented according to the following kinetic Monte Carlo rule. When a particle encounters the membrane, a random number $r_d \in (0,1)$ is generated. If $r_d \leq p(\theta)$, the particle undergoes a scattering collision with the membrane; otherwise, it continues its motion. As foregoing mentioned $p(\theta)$ is the sigmoidal profile given by Eq.~\eqref{eq:13}.

Different membrane configurations were used in this study according to their location in the simulation ($L\times L \times L$) box:
\begin{enumerate}[label=(\roman*)]
	\item Case $W = 0$: absence of thermally controlled membranes in the simulation box.
	\item Case $W = 1$: presence of a single membrane at the center of the box, located at $x_w = 0$ (as seen in Fig.~\ref{fig:01}).
	\item Case $W = 2$: presence of two membranes located at $x_w = -L/2$ and $x_w = L/2$.
	\item Case $W = 3$: presence of three membranes located at $x_w = -L/2$, $x_w = 0$, and $x_w = L/2$.
\end{enumerate}

\section{\label{sec:r} Results and Discussion}

\begin{figure}[!b]
	\centering
	\includegraphics*[scale=0.38,angle=0]{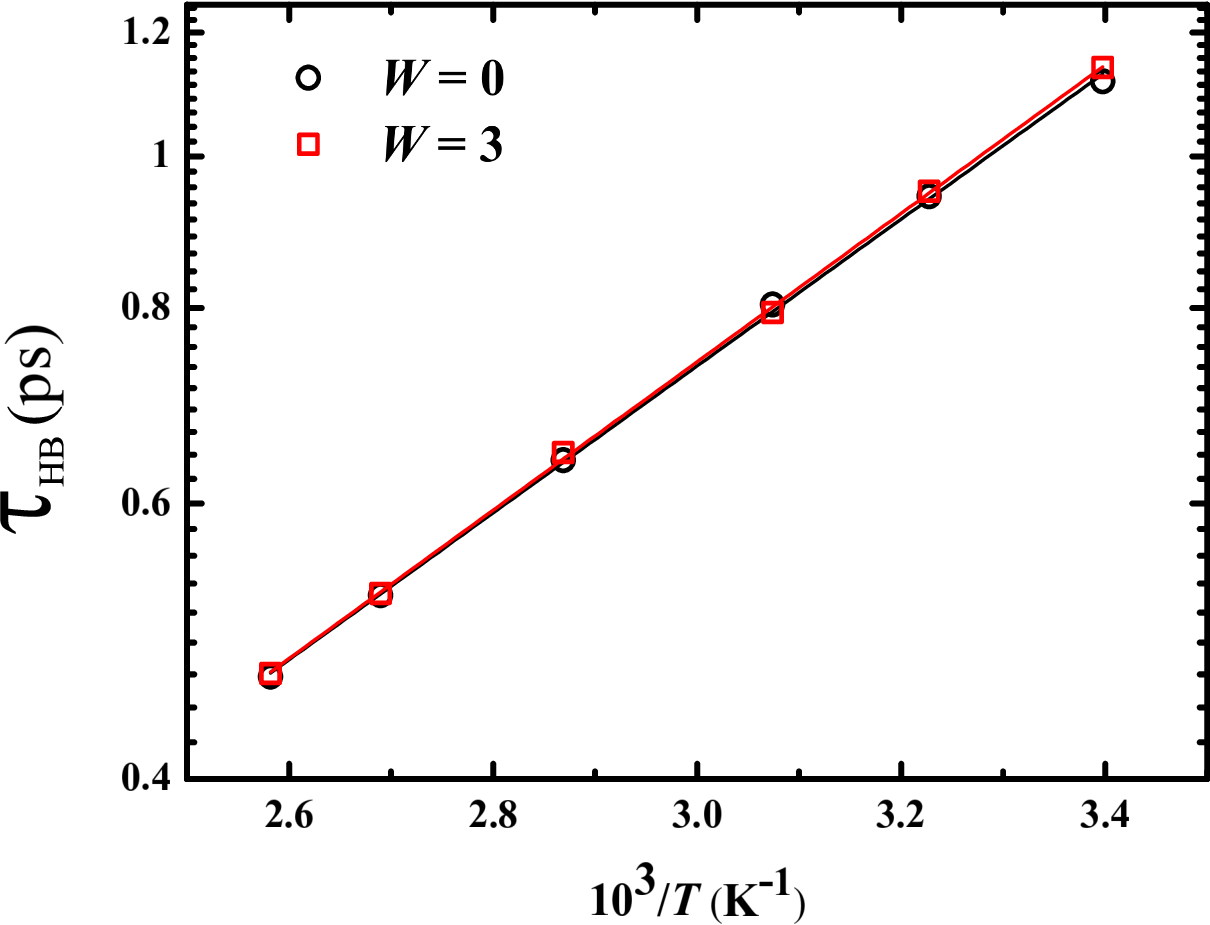}
\caption{Mean hydrogen-bond (HB) lifetime, $\tau_{\mathrm{HB}}$, as a function of the reciprocal temperature for the $W=0$ and $W=3$ cases. The straight lines represent Arrhenius fits to the data. The corresponding pre-exponential factors and activation energies are $\tau_{\mathrm{HB}}^{0}=0.029\,\mathrm{ps}$ and $E_{a}=8.99 \pm 0.10\,\mathrm{kJ\,mol^{-1}}$ for $W=0$, and $\tau_{\mathrm{HB}}^{0}=0.028\,\mathrm{ps}$ and $E_{a}=9.08 \pm 0.08\,\mathrm{kJ\,mol^{-1}}$ for $W=3$.}
\label{fig:07}
\end{figure}

Diffusion calculations were performed over a wide temperature range ($\sim 310\text{--} 465 \, \mathrm{K}$) at two water densities, namely, $0.60\,\mathrm{g/cm^3}$ and $1.0 \,\mathrm{g/cm^3}$, for several TAM configurations. Figures~\ref{fig:03} show the self-diffusion coefficients as a function of temperature for the different TAM configurations and the two water-density regimes. The Arrhenius curves fitted to the data are also shown, and the corresponding activation energies ($E_a$) and pre-exponential factors ($D_0$) are reported in Table~\ref{table:02}. The fitted curves are found to be in reasonable agreement with the Arrhenius law, particularly at high temperatures.

At the lower density ($0.60 \,\mathrm{g/cm^3}$), SPC/E water exhibits a higher activation energy than at the higher density ($1.0 \,\mathrm{g/cm^3}$). Furthermore, the $E_{a}$ values for the different TAM configurations agree with each other within two standard deviations, while exhibiting a slight decrease with increasing number of TAMs in the high-density regime. A similar behavior is observed in the low-density regime, except for the $W=3$ case, for which the TAMs lead to a significant reduction in the system's $E_{a}$. These results suggest that the stochastic action of the TAMs primarily affects the system locally and that a larger number of membranes may be required to effectively reduce the activation energy of the system. Nevertheless, the calculated self-diffusion coefficients generally decrease as the number of TAMs increases, as can be readily inferred from the plot scales in Fig.~\ref{fig:03}. The error bars are comparable in size to the plot symbols and are therefore not shown.

Our $E_a$ values at a density of $1.0 \,\mathrm{g/cm^3}$ are in good agreement with those obtained from depolarized light-scattering experiments ($10.8 \pm 1.0 \,\mathrm{kJ\, mol^{-1}}$) \cite{Conde1983,Conde1984} and previous SPC/E water simulations ($9.3 \pm 1.2 \,\mathrm{kJ\, mol^{-1}}$) \cite{Starr1999}, despite the pressure being allowed to vary in our simulations. The activation energy has been interpreted as the energy required to break a H-bond in liquid water \cite{Conde1983,Conde1984}.

We now turn our attention to the effects of TAMs on the tetrahedrality of water and the HB lifetime. Figure~\ref{fig:04} shows both the tetrahedrality distribution, $P(q)$, and the mean tetrahedrality, $\langle q \rangle$ (see the plot legends), at temperatures $T=310\,\mathrm{K}$ and $T=387\,\mathrm{K}$ for the two $W$ cases considered, at a water density of $1.0,\mathrm{g/cm^3}$. At $T=310\,\mathrm{K}$, we obtained $\langle q \rangle=0.7043 \pm 0.0004$ for the $W=0$ case and $\langle q \rangle=0.7012 \pm 0.0004$ for the $W=1$ case. At $T=387\,\mathrm{K}$, we obtained $\langle q \rangle=0.6327 \pm 0.0006$ for the $W=0$ case and $\langle q \rangle=0.6363 \pm 0.0006$ for the $W=1$ case. As expected, the low-temperature system exhibits a higher degree of tetrahedrality than the high-temperature system for both $W$ cases. Furthermore, the effects of TAMs on $\langle q \rangle$ are noticeable only at the third decimal place, indicating that TAMs have only a weak influence on the overall mean tetrahedrality. Interestingly, this influence can either increase or decrease $\langle q \rangle$, depending on the temperature considered.

To better understand this behavior, it is useful to recall the physical effect of the membranes. According to Eq.~\eqref{eq:13}, at $T=310\,\mathrm{K}$, the probability that a given water molecule undergoes elastic scattering at the membrane is $p\simeq 0.52$, whereas at $T=387\,\mathrm{K}$, this probability increases to $p\simeq 0.75$. Thus, at the lower temperature, the membrane is more likely to allow molecules to pass through without scattering, whereas at the higher temperature, elastic scattering events occur more frequently. Consequently, the perturbation induced by the membrane is stronger at the higher temperature, favoring the local accumulation (or jamming) of water molecules in the vicinity of the membrane.

For a further analysis, we also calculated $\langle q \rangle$ by sampling molecules within narrower virtual boxes centered around the thermally activated membrane, in order to probe the local effects of the TAM on the tetrahedrality of water. Figure~\ref{fig:05} shows $\langle q \rangle$ calculated within virtual boxes centered at the membrane ($x=0$), with different box lengths, at temperatures of $310\,\mathrm{K}$ and $387\,\mathrm{K}$. Two different $W$ cases are considered, with the box length ranging from $L/32$ to $L/2$, being $L=9.8\, \sigma$. The effects of the membrane become more pronounced for box lengths below $2\sigma$, although the differences in $\langle q \rangle$ remain limited to the third decimal place. Once again, these effects can either increase or decrease $\langle q \rangle$, depending on the temperature considered.

Next, we investigate whether the action of the membranes can alter the librational motion of water molecules and, consequently, affect their HB lifetimes. Figure~\ref{fig:06} shows the H-bond lifetime probability distributions at three different temperatures for the $W=0$ case [Fig.~\ref{fig:06a}] and the $W=3$ case [Fig.~\ref{fig:06b}]. As expected, typical exponential tails are observed in all cases. Using Eq.~\eqref{eq:20}, we obtained mean HB lifetimes of $0.943\,\mathrm{fs}$ ($310\,\mathrm{K}$), $0.640\,\mathrm{fs}$ ($349\,\mathrm{K}$), and $0.465\,\mathrm{fs}$ ($387\,\mathrm{K}$) for the $W=0$ case. For the $W=3$ case, the corresponding values are $0.950\,\mathrm{fs}$, $0.647\,\mathrm{fs}$, and $0.467\,\mathrm{fs}$, respectively. These results indicate that TAMs exert only a weak influence on the HB dynamics of water molecules. Since the primary effect of the membranes is to enhance the local jamming of molecules in their vicinity, the presence of TAMs ($W\neq0$) leads, on average, to a slight increase in the HB lifetime compared with the $W=0$ case (absence of membranes). 

The mean HB lifetimes for all cases considered are shown in Fig.~\ref{fig:07}, together with the Arrhenius fits to the data (straight lines). A very small increase in $\tau_{\mathrm{HB}}$ is observed for the $W=3$ case compared with the $W=0$ case at most temperatures. The pre-exponential factor and activation energy are, respectively, $\tau_{\mathrm{HB}}^{0}=0.029\,\mathrm{ps}$ and $E_{a}=8.99 \pm 0.10\,\mathrm{kJ\, mol^{-1}}$ for the $W=0$ case, whereas for the $W=3$ case, we obtained $\tau_{\mathrm{HB}}^{0}=0.028\,\mathrm{ps}$ and $E_{a}=9.08 \pm 0.08,\mathrm{kJ\, mol^{-1}}$. These results further indicate that TAMs exert only a localized and very weak influence on the HB dynamics of water molecules. The obtained activation energies are in very good agreement with those reported in previous studies of SPC/E water: $8.8 \pm 1.0\,\mathrm{kJ\, mol^{-1}}$ using an energetic HB definition \cite{Starr2000} and $9.3 \pm 1.2\,\mathrm{kJ\, mol^{-1}}$ using a geometric HB definition \cite{Starr1999}. Moreover, our results lie within the range of experimental values obtained from quasi-elastic coherent neutron scattering, $E_{a}=7.7\,\mathrm{kJ\, mol^{-1}}$ \cite{Teixeira2006}, and depolarized light-scattering measurements, $E_{a}=10.8\,\mathrm{kJ\, mol^{-1}}$ \cite{Conde1983,Conde1984}.

\section{Conclusions}

In this work, we investigated the effects of thermally activated membranes (TAMs) on the dynamical and structural properties of SPC/E water over a broad high-temperature range using molecular dynamics simulations. The membranes were modeled as stochastic interfaces that induce elastic scattering events according to a temperature-dependent sigmoidal probability.

Our results show that TAMs primarily affect the translational dynamics of water. In general, the self-diffusion coefficient decreases as the number of membranes increases, indicating that membrane-induced scattering hinders molecular transport. Nevertheless, the temperature dependence of the diffusion coefficient remains reasonably well described by an Arrhenius-like behavior over the temperature range investigated. The effective activation energy exhibits only moderate variations among the different membrane configurations, suggesting that the membranes modify molecular mobility without substantially altering the underlying thermally activated diffusion mechanism.

In contrast, the tetrahedral structure and hydrogen-bond dynamics are only weakly affected by the membranes. The mean tetrahedral order parameter, $\langle q\rangle$, shows small variations in the presence of TAMs, with more noticeable changes restricted to regions in the immediate vicinity of the membranes. Similarly, the mean HB lifetime is only slightly modified by membrane-induced scattering. These results indicate that, while TAMs can significantly influence long-range translational transport, their effects on the local structure and hydrogen-bond dynamics of water remain comparatively weak and spatially localized.

Finally, the present results demonstrate that thermally activated stochastic interfaces provide a mechanism for controlling molecular transport while largely preserving the local structural properties of liquid water. The TAM model therefore offers a simple framework for exploring thermally controlled transport in confined and membrane-mediated molecular systems.

\section{Acknowledgements}

We wish to thank UFERSA for computational support.




\bibliographystyle{model3a-num-names}
%%\bibliography{<your-bib-database>}
%\bibliography{gaspack}

\end{document}